\documentclass[preprint,aps,amsmath,amssymb,pra,showpacs,showkeys]{revtex4-1}
\usepackage{graphicx}

\begin{document}
\title{The influence of the electric polarization of hydrogen atoms\\on the red shift of its spectral lines}
\author{V. S. Severin}
\affiliation{Department of General and Applied Physics, National Aviation University, Cosmonaut Komarov prospect 1, Kyiv 03058, Ukraine}
\email{severinvs@ukr.net}
\date{\today}

\begin{abstract}
The Lorentz oscillator system is studied to interpret the spectral lines of hydrogen atoms. The dielectric
constant of this system is analyzed, which takes into account the electrical polarization of hydrogen atoms.

This dielectric constant gives the red shift of the spectral line and the appearance of the optical spectrum
dip. This dip is on the blue side of the spectral position of the shifted line.

The value of this red shift and the width of this dip strongly depend on the hydrogen atom concentration 
and the spectral position of the not shifted line. This red shift increases with an increase in the hydrogen 
atom concentration.
\end{abstract}



\maketitle

\section{\label{sec1}Introduction}
The substance spectral line research is the subject of optical spectroscopy since its origin. Modern
physics and astrophysics use it \cite{Bakhshiev1987,Sobolev1985,Pradhan2011,Rybicki2004,Roemer2005}. 
The Lorentz oscillator gives a simple but successful classical model of the spectral line. It is widely
utilized to interpret substance spectral lines \cite{Pradhan2011,Rybicki2004,Roemer2005}.

The dielectric constant of a substance expressed through this substance electrical conductivity
gives optical properties of this substance \cite{Roemer2005,Platzmanand1973,Il’inskiy1989,
Landau1982}. However, both the external electrical conductivity of this substance and its internal 
electrical conductivity characterize this substance \cite{Kubo1966,Izuyma1961,Eykholt1986}.

According to the definition of the dielectric constant of a substance, the internal electrical conductivity
of this substance determines this constant \cite{Platzmanand1973,Il’inskiy1989,Landau1982,
Kubo1966,Izuyma1961,Eykholt1986}.

The external electrical conductivity of a substance is used traditionally instead of its internal electrical
conductivity in the dielectric constant of this substance. This use is the approximation that does not take 
into account the electrical polarization of charge carriers of this substance \cite{Severin1990,
Severin1994,Vakulenko2007,Vakulenko2007a,Severin2009,Severin2012,Severin2017}. The difference between the 
external electromagnetic field acting on this substance and the internal electromagnetic field acting in this 
substance is absent in this approximation.

Optical transitions of electrical charges in a substance take place under the influence of the 
internal electromagnetic field in this substance. Molecular spectroscopy considers the effect of the 
difference between external and internal electromagnetic fields in a substance. Molecular spectroscopy 
shows the single-molecule spectrum that can differ from the molecule system spectrum \cite{Bakhshiev1987}.

The article is organized as follows: Sec.~\ref{sec2} discusses the difference between external and internal 
electrical conductivities. Sec.~\ref{sec3} gives optical properties of the substance modeled by the  
Lorentz oscillator system that does take into account the electrical polarization of this system. 
Sec.~\ref{sec4} considers the application of the results of Sec.~\ref{sec3} to the spectral lines of 
hydrogen atoms. Sec.~\ref{sec3} gives the final discussion and conclusions. This article uses Gaussian units.

\section{\label{sec2}External and internal electrical conductivities}
The article considers an optically isotropic substance. Let the external electrical conductivity and the 
internal electrical conductivity of this substance be scalars but not tensors. They depend on 
the frequency $\omega$ of the light wave but do not depend on its wave vector. Hence, the dependence 
of the external electric field and the internal electric field on the wave vector not considered.

Let the external electric field ${{\mathbf{d}}_{t}}=\mathbf{D}\left(\omega\right)\exp(i\omega t)$ 
act on the substance. Here, $\textit{t}$ is time; $\mathbf{D}\left(\omega\right)$ is the this field 
amplitude. The electric current density that has the amplitude $\mathbf{j}(\omega)$ arises in the 
substance under this field action. The following equation gives this current density
\begin{equation}
\label{eqA1}
\mathbf{j}\left(\omega\right)=\sigma(\omega)\mathbf{E}\left(\omega\right)=s(\omega)\mathbf{D}
\left(\omega\right) .
\end{equation}
Here, $\mathbf{E}\left(\omega\right)$ is the internal electric field that acts in the substance.

The relation~(\ref{eqA1}) defines two different conductivities \cite{Kubo1966, Izuyma1961,Eykholt1986}. 
The external conductivity $s\left(\omega\right)$ is determined by the Kubo formula \cite{Kubo1966}. 
It differs from the internal conductivity $\sigma\left(\omega\right)$. 
This difference between $s\left(\omega\right)$ and $\sigma\left(\omega\right)$ is due to the difference 
between the fields $\mathbf{E}\left(\omega\right)$ and $\mathbf{D}\left(\omega\right)$ because the 
following equation takes place 
\begin{equation}
\label{eqA2}
\mathbf{D}\left(\omega\right)=\mathbf{E}\left(\omega\right)+4\pi\mathbf{P}\left(\omega\right).
\end{equation}
Here, $\mathbf{P}(\omega)$ is the substance electrical polarization vector as well as \cite{Roemer2005,
Platzmanand1973,Il’inskiy1989,Landau1982,Kubo1966,Izuyma1961,Eykholt1986}
\begin{equation}
\label{eqA3}
\mathbf{P}(\omega)=\frac{\mathbf{j}(\omega)}{i\omega}.
\end{equation}

The following relationship determines the substance dielectric constant $\epsilon\left(\omega\right)$ 
\cite{Roemer2005,Platzmanand1973,Il’inskiy1989,Landau1982,Kubo1966,Izuyma1961,Eykholt1986}
\begin{equation}
\label{eqA4}
\mathbf{D}(\omega)=\epsilon (\omega)\mathbf{E}(\omega) .
\end{equation}
Eqs.~(\ref{eqA1})--(\ref{eqA4}) give
\begin{equation}
\label{eqA5}
\epsilon(\omega)=1+\frac{4\pi}{i\omega}\sigma(\omega).
\end{equation}
Further, we have the following from Eqs.~(\ref{eqA1}),~(\ref{eqA4})
\begin{equation}
\label{eqA6}
\sigma(\omega)=s(\omega)\epsilon(\omega).
\end{equation}
In addition, Eqs.~(\ref{eqA5}),~(\ref{eqA6}) give the following relationship between $\textit{s}$ 
and $\sigma$ 
\begin{equation}
\label{eqA7}
\sigma (\omega)=\frac{1}{1-4\pi{s(\omega)}/{(i\omega)}\;}s(\omega).
\end{equation}

The conductivities $\textit{s}$ and $\sigma$ of the Lorentz oscillator system were 
considered in ~\cite{Severin2009}. 

In general, $\textit{s}$ and $\sigma$ are not scalar quantities, and there are complex matrix 
relations instead of Eqs.~(\ref{eqA1})--(\ref{eqA7}).

\section{\label{sec3} Electrical polarization influence on spectral lines} 
Let us consider the system of identical Lorentz oscillators having the concentration \textit{n}. 
Let \textit{m} be the oscillator mass, \textit{e} be the oscillator charge, and $\gamma$ 
be the damping oscillator factor. Lat $\omega_0$ be the frequency of one oscillator before taking 
into account the electrical polarization of the oscillator system. The frequencies of these 
oscillators are associated with the spectral lines of atoms in this article.

The dielectric constant $\epsilon(\omega)$ of this Lorentz oscillator system that does take 
into account the electric polarization has the form of~\cite{Severin2009}
\begin{equation}
\epsilon(\omega)=1+\omega_p^2\frac{1}{{\Omega_0^2-{\omega^2}+i\gamma\omega}}
\label{eq1}.
\end{equation}
Here, the following formulas give the frequencies $\omega_{p}$ and $\Omega_0$  
\begin{equation}
\omega_{p}^{2}=\frac{4\pi{{e}^{2}}n}{m}\
\label{eq2},
\end{equation}
\begin{equation}
\Omega_0^2=\omega_0^2\left({1-{{\left({\frac{{{\omega_p}}}{{{\omega_0}}}}\right)}^2}}\right)
\label{eq3}.
\end{equation}
The frequency $\Omega_0$ is the frequency of one oscillator after taking into 
account the electrical polarization of this oscillator system \cite{Severin2009}.

The dielectric constant of this system of Lorentz oscillators ${{\epsilon}_{s}}(\omega)$ 
that does not take into account the electric polarization has the form 
of \cite{Severin2009}
\begin{equation}
{{\epsilon}_{s}}(\omega)=1+\omega_{p}^{2}\frac{1}{\omega_{0}^{2}-{{\omega}^{2}}+i\gamma\omega}
\label{eq4}.
\end{equation}

Eq.~(\ref{eq3}) gives the decrease in the frequency $\Omega_0$ when the oscillator 
concentration \textit{n} increases. There is an agreement between this theoretical result 
and the experimental results of the optical spectra of lattice vibrations \cite{Severin2009}.

Let us consider the spectral line of a substance atom. We believe that Eqs.~(\ref{eq1})--(\ref{eq4})
give the dielectric constant of this substance, in which \textit{n} is the atom concentration, 
\textit{e} is the electron charge, and \textit{m} is the electron mass. In this case, $\omega_{0}$ is 
the spectral line frequency of an isolated atom or the oscillator frequency without taking 
into account the electric polarization of the oscillator system. However, $\Omega_0$ is the 
spectral line frequency of a substance atom or the oscillator frequency with taking into account 
this electrical polarization.

The following formulas give the real and the imaginary part of the dielectric constants 
$\epsilon(\omega)$ and ${{\epsilon}_{s}}(\omega)$ 
\begin{equation}
{{\epsilon}_{1}}\left(\omega\right)\equiv\operatorname{Re}\epsilon\left(\omega\right)=\frac{
\left(\omega_{0}^{2}-{{\omega}^{2}}\right)\left(\Omega_{0}^{2}-{{\omega}^{2}}\right)+{{\gamma}^{2}}
{{\omega }^{2}}}{{{\left(\Omega_{0}^{2}-{{\omega}^{2}}\right)}^{2}}+{{\gamma}^{2}}{{\omega}^{2}}} ,
\label{eq5}
\end{equation}
\begin{equation}
{{\epsilon}_{2}}\left(\omega\right)\equiv-\operatorname{Im}\epsilon\left(\omega\right)=\omega_{p}^{2}
\frac{\gamma\omega}{{{\left(\Omega_{0}^{2}-{{\omega}^{2}}\right)}^{2}}+{{\gamma}^{2}}{{\omega}^{2}}} ,
\label{eq6}
\end{equation}
\begin{equation}
{{\epsilon}_{s1}}\left(\omega\right)\equiv\operatorname{Re}{{\epsilon}_{s}}
\left(\omega\right)=\frac{\left(\omega_{0}^{2}-{{\omega}^{2}}\right)\left(\omega_{0}^{2}
-{{\omega}^{2}}+\omega_{p}^{2}\right)+{{\gamma}^{2}}{{\omega }^{2}}}{{{\left(\omega_{0}^{2}
-{{\omega}^{2}}\right)}^{2}}+{{\gamma}^{2}}{{\omega}^{2}}},
\label{eq7}
\end{equation}
\begin{equation}
{{\epsilon}_{s2}}\left(\omega\right)\equiv-\operatorname{Im}{{\epsilon}_{s}}
\left(\omega\right)=\omega_{p}^{2}\frac{\gamma\omega}{{{\left(\omega_{0}^{2}-{{\omega}^{2}}
\right)}^{2}}+{{\gamma}^{2}}{{\omega}^{2}}} .  
\label{eq8}
\end{equation}
       
Light waves do not pass through a substance in which the real part of the dielectric constant 
of this substance is negative at the frequency of these waves. Therefore, the optical spectrum 
of this substance has a dip at these wave frequencies. If there is the approximation $\gamma\to0$, 
then Eq.~(\ref{eq5}) gives that the condition ${{\epsilon}_{1}}\left(\omega\right)<0$	takes 
place at the frequency range 
\begin{equation}
{{\Omega}_{0}}=\sqrt{\omega_{0}^{2}-\omega_{p}^{2}}<\omega<{{\omega}_{0}} .   
\label{eq9}
\end{equation}
 
However, in the same approximation, Eq.~(\ref{eq7}) gives that the condition 
${{\epsilon}_{s1}}\left(\omega\right)<0$ takes place at the frequency range 		
\begin{equation}
{{\omega}_{0}}<\omega<\sqrt{\omega_{0}^{2}+\omega_{p}^{2}} .   
\label{eq10}
\end{equation}
		
Eqs.~(\ref{eq9}) and~(\ref{eq10}) give the fact that the model of the Lorentz oscillator 
system that takes the electrical polarization of this system into account and the model of 
the Lorentz oscillator system that not takes the electrical polarization of this system into 
account have the dip in optical spectrum in different frequency ranges.	
		
The maximum of the imaginary part of the dielectric constant determines the spectral line place  
if the frequency dependence of the real part of this dielectric constant not taken 
into account. This article uses such a method. In this case, formulas~(\ref{eq6}) and~(\ref{eq8}) 
give different places of the spectral line.		
		
The wavelength of light ${{\lambda}_{0}}=c2\pi/{{\omega}_{0}}$ corresponds to 
the Lorentz oscillator system without taking into account the polarization. This wavelength 
${{\lambda}_{0}}$ gives the spectral line of an isolated atom in the optical spectrum.  Here, 
$\textit{c}$ is the light velocity in free space. 

The wavelength of light ${{\lambda}_{1}}=c2\pi/{{\Omega}_{0}}$ corresponds to the Lorentz 
oscillator system taking into account the electrical polarization. It gives  the spectral line 
of a substance atom. This polarization gives the next wavelength shift to the red side 
${\Delta\lambda}={{\lambda}_{1}}-{{\lambda}_{0}}$. Besides, following Eq. (9), the dip in the 
optical spectrum is on the blue side relative to the place of this shifted line.
		
The redshift parameter of the spectral line is $z({{\lambda}_{0}})=\Delta{\lambda}/{{\lambda}_{0}}$. 
Eq.~(\ref{eq3}) gives		
\begin{equation}
z\left({{\lambda}_{0}}\right)=\frac{{{\lambda}_{1}}-{{\lambda}_{0}}}{{{\lambda}_{0}}}
=\frac{{{\omega}_{0}}}{{{\Omega}_{0}}}-1=\frac{1}{\sqrt{1-{\omega_{p}^{2}}/{\omega_{0}^{2}}\;}}-1 .   
\label{eq11}
\end{equation}
Let be ${{\lambda}_{p}}=c2\pi/{{\omega}_{p}}=\left({c}/{e}\;\right)\sqrt{{\pi m}/{n}\;}$. Then, 
Eq.~(\ref{eq11}) gives 
\begin{equation}
z\left({{\lambda}_{0}}\right)=\frac{1}{\sqrt{1-{\lambda_{0}^{2}}/{\lambda_{p}^{2}}\;}}-1
=\frac{1}{\sqrt{1-{n}/{N({{\lambda}_{0}})}\;}}-1 .   
\label{eq12}
\end{equation}
Here,
\begin{equation}
N({{\lambda}_{0}})=\frac{\pi m{{c}^{2}}}{\lambda_{0}^{2}{{e}^{2}}}=\frac{\pi}{{{r}_{0}}\lambda_{0}^{2}}   
\label{eq13}
\end{equation}
is the characteristic concentration that depends on the wavelength ${{\lambda}_{0}}$ and the parameter
${{r}_{0}}={{{e}^{2}}}/{\left(m{{c}^{2}}\right)}$. If \textit{m} be the electron mass and \textit{e} be 
the electron charge, this parameter is the classical electron radius.
									
According to Eq.~(\ref{eq11}), only if the condition ${{\omega}_{0}}>{{\omega}_{p}}$ takes place, 
the line having the wavelength ${\lambda_{1}}={{\lambda}_{0}}+\Delta\lambda$ is observed in the spectrum. 
Otherwise, the limit ${{\omega}_{0}}\to{{\omega}_{p}}$ gives $\Delta\lambda\to\infty$. The vibrations 
of the oscillator disappear in this limit. In that case, the spectral line vanishes. 

Namely, the observed wavelength ${\lambda_{1}}={{\lambda}_{0}}+\Delta\lambda$ of the spectral line of  
the substance is present in its spectrum only if the initial wavelength ${{\lambda}_{0}}$ 
is less than the wavelength ${{\lambda}_{p}}$. Therefore, the following condition must 
take place to observe this line
\begin{equation}
{{\lambda}_{0}}<{{\lambda}_{p}}=\sqrt{\frac{\pi}{{{r}_{0}}n}}=\frac{c}{e}\sqrt{\frac{\pi m}{n}}.    
\label{eq14}
\end{equation}

Condition~(\ref{eq14}) gives the following restriction on the atom concentration \textit{n}
\begin{equation}
n<N({{\lambda}_{0}}).
\label{eq15}
\end{equation}
Condition~(\ref{eq15}) indicates that the observed wavelength ${\lambda_{1}}={{\lambda}_{0}}+\Delta\lambda$ 
of the spectral line of the substance exists in its spectrum only if the concentration 
\textit{n} is less than the concentration $N({{\lambda}_{0}})$, which depends on the wavelength 
${{\lambda}_{0}}$.
		
\section{\label{sec4}Spectral lines of hydrogen atoms}
Let us consider hydrogen atom spectral lines in terms of the results of the previous section.

Hydrogen exists in the molecular state under usual laboratory conditions. A small part of its 
molecules are destroyed to atoms (for example, by an electric discharge) to observe the 
spectrum of hydrogen atoms. However, the dependence of the spectral place of the hydrogen atom 
line on the hydrogen atom concentration is not studied. On the other hand, the interstellar 
medium has hydrogen atoms having different concentrations. There are many experimental 
results of their optical spectra.

As stated above, the wavelength of the spectral line of hydrogen atoms ${\lambda_{1}}$ 
depends strongly on the atom concentration \textit{n} and the characteristic 
concentration $N({{\lambda}_{0}})$ determined by the spectral line wavelength of an 
isolated atom ${\lambda}_{0}$.

\begin{figure}[!ht]
\includegraphics{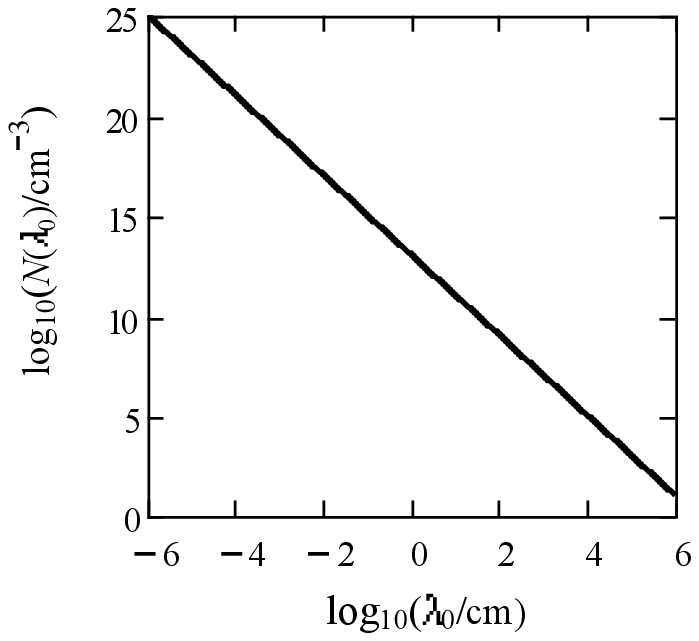}
\caption{\label{fig1} The dependence of the concentration $N({{\lambda}_{0}})$ on the wavelength 
${\lambda}_{0}$.}
\end{figure}

Figure~\ref{fig1} shows the dependence of $N({{\lambda}_{0}})$ on the wavelength 
${\lambda}_{0}$. It gives that the value of $N({{\lambda}_{0}})$ is very high in the optical range 
of the wavelengths ${\lambda}_{0}$. However, the value of the red shift parameter of 
the rarefied gas of hydrogen atoms is very little in this wave range. 
The interstellar medium has the hydrogen atom concentration $\textit{n}$ 
in the range of $n={{10}^{-3}}-{{10}^{4}}~\text{c}{{\text{m}}^{-3}}$ (Table 1.3
~\cite{Draine2011}). At this atom concentration, the red shift of spectral line may be 
significant in the radio-frequency range of wavelengths. For instance, the concentration 
$N({{\lambda}_{0}})$ is about ${{10}^{9}}~\text{c}{{\text{m}}^{-3}}$ at the wavelength 
${{\lambda}_{0}={10}^{2}}~\text{c}{{\text{m}}}$.

\begin{figure}[!ht]
\includegraphics{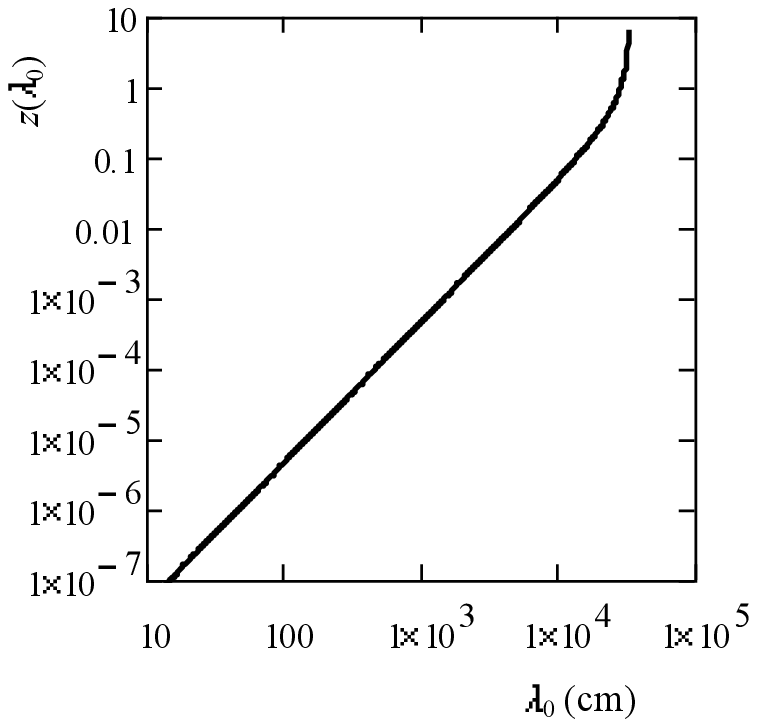}
\caption{\label{fig2} The dependence of the red shift parameter $z({{\lambda}_{0}})$ on the wavelength 
${{\lambda}_{0}}$ at the concentration $n={{10}^{4}}~\text{c}{{\text{m}}^{-3}}$.}
\end{figure}

Figure~\ref{fig2} shows the dependence of the red shift parameter $z({{\lambda}_{0}})$ 
on the wavelength ${{\lambda}_{0}}$ at the concentration $n={{10}^{4}}~\text{c}{{\text{m}}^{-3}}$. 
It gives that the red shift parameter $z({{\lambda}_{0}})$ strongly depends on the wavelength 
${{\lambda}_{0}}$.

The interstellar matter exists both in the gas state and in the dust state~\cite{Sobolev1985,
Draine2011,Dyson1997,Boren1986}. Hydrogen is the main component of the interstellar 
matter~\cite{Sobolev1985,Draine2011}. Because of the low temperature of interstellar 
medium, the dust surface may be covered by frozen hydrogen~\cite{Draine2011,Dyson1997,Boren1986}. 
Hydrogen atoms can penetrate some solids and reach a high atom concentration that 
comparable to the atom concentration of solids~\cite{Gamburg1989,Fukai2005}. 
However, at the same time, they can remain in the atomic state ~\cite{Gamburg1989,Fukai2005}.

Dust clouds are of great importance for the physical processes in quasars~\cite{Fan2006}. 
Possibly, depending on the conditions of the dust particles, the hydrogen atom concentration 
in these particles can vary from small values to values that are inherent in solids.

\begin{figure}[!ht]
\includegraphics{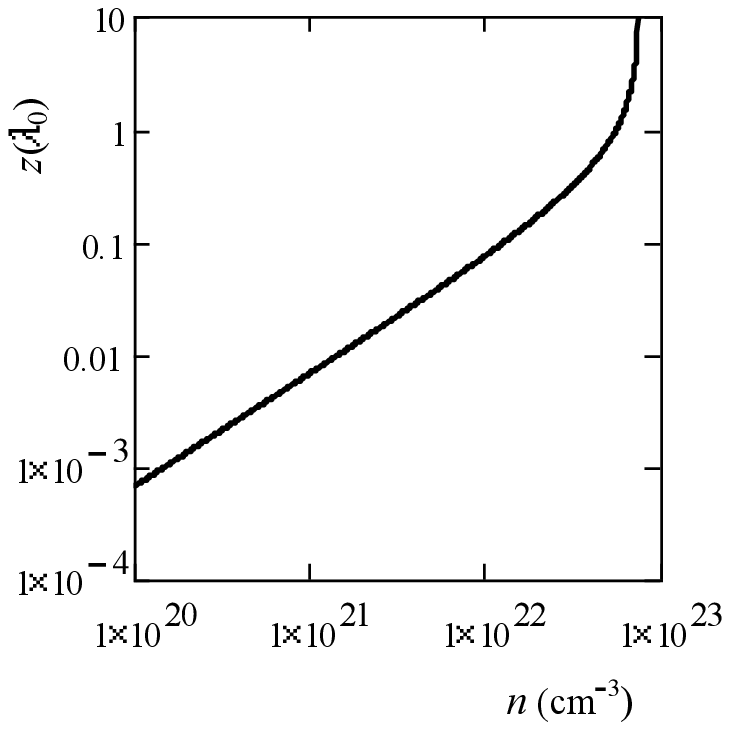}
\caption{\label{fig3} The dependence of the red shift parameter $z({{\lambda}_{0}})$ on the 
concentration $n$ at the wavelength ${\lambda}_{0}~=~{1.216\cdot{{10}^{-5}}~\text{cm}}$.}
\end{figure}

The Ly$\alphaα$ line of optical spectrum of an isolated hydrogen atom has the wavelength 
${\lambda}_{0}~=~1.216\cdot{{10}^{-5}}~\text{cm}$. Eq.~(\ref{eq13}) gives the value 
$N\left({{\lambda}_{0}}\right)=7.54\cdot{{10}^{22}}~\text{c}{{\text{m}}^{-3}}$
at this value of ${\lambda}_{0}$. Such a value of $N({\lambda}_{0})$ is close to the 
atom concentration in a solid. Solid bodies have the atom concentration of the order of 
${{10}^{23}}~\text{c}{{\text{m}}^{-3}}$. Therefore, Eq.~(\ref{eq12}) gives 
$z\left({{\lambda}_{0}}\right)>>1$ at the hydrogen atom concentration $n$ in dust 
particles that are close to this value of $N({\lambda}_{0})$. Figure~\ref{fig3} shows the 
dependence of the redshift parameter $z({{\lambda}_{0}})$ on the hydrogen 
atom concentration $n$ at the wavelength ${\lambda}_{0}=1.216\cdot{{10}^{-5}}~\text{cm}$. It 
gives that a quasar dust cloud can give a high value of this parameter if the dust particles 
of this cloud have a high hydrogen atom concentration.

\section{\label{sec5}Discussion and conclusions}
If we take into account the electrical polarization of hydrogen atoms, then the red shift 
of spectral lines of these atoms and the dip in their optical spectrum appear. This dip is 
on the blue side relative to the place of the shifted line. The red shift value and the width 
of the dip strongly depend on the hydrogen atom concentration and the  place of the not shifted 
line.

This effect is significant for interstellar hydrogen atoms if the wavelength of the spectral 
lines of these atoms is in the radio frequency range.

The value of this effect can be significant for the optical spectrum of hydrogen atoms that 
can be in interstellar dust clouds. Perhaps this is due to the high concentration of hydrogen 
atoms in the dust substance.

Quasars in which dust clouds are of significant importance~\cite{Fan2006} give experimental 
spectra with such features~\cite{Pradhan2011,Fan2006,Fan2012}. The line Ly$\alphaα$ of hydrogen 
atoms of quasars has a high red shift (Fig. 14.4~\cite{Pradhan2011}, Fig. 1~\cite{Fan2006}, 
and Fig. 8~\cite{Fan2012}). Besides, this line has a dip on the blue side relative to the 
shifted line (Fig. 14.4~\cite{Pradhan2011}, Fig. 1~\cite{Fan2006}, and Fig. 8~\cite{Fan2012}).

These results are essential for astrophysics. Therefore, it is desirable to test them experimentally 
in terrestrial laboratories. Indeed, in laboratory conditions, there are methods to obtain and 
stabilize hydrogen atoms both in the state of free gas~\cite{Leonas1981, SiIvera1982, Gamburg1989} 
and as part of a solid~\cite{Gamburg1989,Fukai2005}.

\bibliography{Severin2020redshift}

\end{document}